\documentclass[twocolumn,showpacs,preprintnumbers,amsmath,amssymb]{revtex4}

\usepackage{graphicx}
\usepackage{dcolumn}
\usepackage{bm}
\usepackage{dsfont}
\usepackage{mathrsfs}
\usepackage{amsmath}				

\newcommand{\be}{\begin{equation}}
\newcommand{\ee}{\end{equation}}
\newcommand{\bey}{\begin{eqnarray}}
\newcommand{\eey}{\end{eqnarray}}

\begin{document} 
\title{Winding expansion techniques for lattice QCD with chemical potential}
\author{Julia Danzer} 
\author{Christof Gattringer}
\affiliation{
\vspace{3mm}Institut f\"ur Physik, FB Theoretische Physik,
Universit\"at Graz
\vskip0mm 8010 Graz, Austria
\vspace{4mm}
}

\date{September 16, 2008}

\begin{abstract}
\vspace{4mm}
We analytically derive a decomposition of the lattice fermion determinant for
Wilson's Dirac operator with chemical potential into winding sectors, i.e.,
factors with a fixed number of quarks. Dividing the lattice into four domains,
the determinant is factorized into terms which can be classified with respect to
the winding number of the closed loops they consist of. The individual factors
are expressed in terms of subdeterminants and propagators on the domains of the
lattice. We numerically analyze properties of the factorization formula and
discuss two applications for the determination of canonical partition functions
with a fixed quark number: A speedup for the Fourier transformation technique
through a dimensional reduction, and a power series expansion.   
\end{abstract}


\maketitle

\section{Introduction} 
The lattice formulation provides a fully gauge invariant non-perturbative
approach to QCD. An important aspect of the lattice discretization is the
possibility of using numerical Monte Carlo techniques to obtain non-perturbative
results. An application where the success so far was rather limited is lattice
QCD with chemical potential. The reason is the sign problem of the fermion
determinant with a chemical potential, which forbids a direct application of the
Monte Carlo method such that one is reduced to reweighting methods. 

In these attempts the use of canonical partition functions with a fixed quark 
(or baryon) number has started to play an important role recently
\cite{hasenfratz}--\cite{Forcrand}. For their use it is necessary to have an
efficient method for evaluating the canonical partition function at a fixed
quark number. This problem is in turn related to organizing the fermion
determinant, which may be viewed as a collection of closed loops, with respect
to the winding number of these loops around compactified time. 

In this article we present an explicit decomposition of the fermion determinant
for Wilson's Dirac operator into sectors with definite winding numbers. The
approach is based on a partition of the lattice into four domains, and the
contributions to the individual winding sectors may be written in terms of
propagators on these lattice domains. The terms in in our factorization formula
for the fermion determinant are analyzed numerically. We discuss two possible
applications of our results to the evaluation of the canonical partition
function at fixed quark number.

\section{Partition of the lattice into four domains}

\begin{figure}[t]
\begin{center}
\includegraphics[width=80mm,clip]{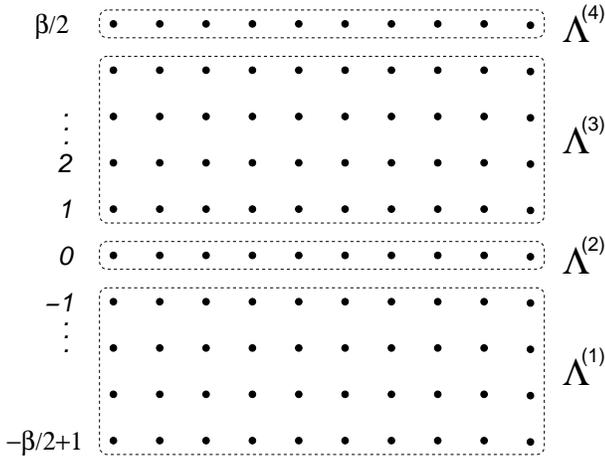} 
\end{center}
\caption{Partition of the lattice into four domains 
$\Lambda^{(i)}, i = 1,2,3,4$: The vertical direction in
the plot is time, while the dots in horizontal direction represent a whole time
slice. On the left-hand side we indicate the values of the time argument. The
lattice sites in the same domain are enclosed in a dashed contour, and we label
the domains on the right-hand side. \hfill} \label{latticeplot} \end{figure}

The starting point of our approach is a partition of a $L^3 \times \beta$ 
lattice $\Lambda$ into four domains with different ranges of the time component
$x_4$:
\begin{eqnarray}
\Lambda^{(1)} & = & \{ (\vec{x},x_4) \, | \; \vec{x} \in \Lambda_s, \,
x_4 = -\beta/2 + 1, \; ... \; -1 \}  \, ,
\nonumber \\
\Lambda^{(2)} & = & \{ (\vec{x},x_4) \, | \; \vec{x} \in \Lambda_s, \,
x_4 = 0 \}  \, ,
\nonumber \\
\Lambda^{(3)} & = & \{ (\vec{x},x_4) \, | \; \vec{x} \in \Lambda_s, \,
x_4 = 1, \; ... \; \beta/2 - 1 \}  \, ,
\nonumber \\
\Lambda^{(4)} & = & \{ (\vec{x},x_4) \, | \; \vec{x} \in \Lambda_s, \,
x_4 = \beta/2 \}  \, .
\label{latticedomains}
\end{eqnarray}
We stress that $\beta$ is a positive integer, the inverse temperature in lattice
units. $\Lambda_s$ denotes the spatial part of the lattice. The full lattice
$\Lambda$ is the union of the four sub-lattices $\Lambda^{(i)},\, i = 1,2,3,4$.
The partition of the lattice is illustrated in Fig.~\ref{latticeplot}.

Wilson's lattice Dirac operator $D(x,y)$ which we use here can be decomposed
into pieces $D^{(i)}$ acting in an individual domain $\Lambda^{(i)}$ and terms
$D^{(i,j)}$ that connect the neighboring domains $\Lambda^{(i)}, \Lambda^{(j)},
i \neq j$. For that purpose we need the operators
\begin{eqnarray}
D^{(i)}(x,y) \; & : & x,y \in \Lambda^{(i)} \; , \; i = 1,2,3,4 ,
\label{subdiracs} \\
D^{(1,2)}(x,y) & : & \vec{x},\vec{y} \in \Lambda_s \; , \;
x_4 = -1, y_4 = 0 ,
\nonumber \\
D^{(2,1)}(x,y) & : & \vec{x},\vec{y} \in \Lambda_s \; , \;
x_4 = 0, y_4 = -1 ,
\nonumber \\
D^{(2,3)}(x,y) & : & \vec{x},\vec{y} \in \Lambda_s \; , \;
x_4 = 0, y_4 = 1 ,
\nonumber \\
D^{(3,2)}(x,y) & : & \vec{x},\vec{y} \in \Lambda_s \; , \;
x_4 = 1, y_4 = 0 ,
\nonumber \\
D^{(3,4)}(x,y) & : & \vec{x},\vec{y} \in \Lambda_s \; , \;
x_4 = \beta/2-1, y_4 = \beta/2 ,
\nonumber \\
D^{(4,3)}(x,y) & : & \vec{x},\vec{y} \in \Lambda_s \; , \;
x_4 = \beta/2, y_4 = \beta/2-1 ,
\nonumber \\
D^{(4,1)}(x,y) & : & \vec{x},\vec{y} \in \Lambda_s \; , \;
x_4 = \beta/2, y_4 = -\beta/2 + 1 ,
\nonumber \\
D^{(1,4)}(x,y) & : & \vec{x},\vec{y} \in \Lambda_s \; , \;
x_4 = -\beta/2 + 1, y_4 = \beta/2 ,
\nonumber
\end{eqnarray}
where we also display which values the arguments $x$ and $y$ may assume. The
term $D^{(1)}$ is given by (we set the lattice constant to 1 throughout this
paper and all numerical results are given in lattice units) 
\begin{eqnarray}
\hspace*{-7mm} && 
D^{(1)}(x,y) \, = \, \delta_{\vec{x},\vec{y}} \, \delta_{x_4,y_4}  
\label{D11} \\
\hspace*{-7mm} && \;\;\; 
- \, \kappa \sum_{j = \pm 1}^{\pm 3} 
\frac{ 1 \mp \gamma_{|j|}}{2} \, U_j(\vec{x},x_4) \, \delta_{\vec{x} + \hat{j},\vec{y}} \, \delta_{x_4,y_4} \,
\nonumber \\
\hspace*{-7mm} && \;\;\; 
- \, \kappa \!\!\!\!\!\!\! \sum_{n_4 = -\frac{\beta}{2}+1}^{-2} \!
\frac{ 1 \! - \! \gamma_4}{2} \, U_4(\vec{x},n_4) \, \delta_{\vec{x},\vec{y}} \, \delta_{x_4,n_4} \, \delta_{y_4,n_4+1}
\nonumber \\
\hspace*{-7mm} && \;\;\; 
- \, \kappa \!\!\!\!\!\!\! \sum_{n_4 = -\frac{\beta}{2}+2}^{-1} \!
\frac{ 1 \! + \! \gamma_4}{2} \, U_4(\vec{x},n_4\!-\!1)^\dagger \, 
\delta_{\vec{x},\vec{y}} \, \delta_{x_4,n_4} \, \delta_{y_4,n_4-1} \, ,
\nonumber 
\end{eqnarray}
where we have defined the hopping parameter $\kappa = 1/(4+m)$ with $m$ being
the bare quark mass \footnote{We remark that our definition of the hopping
parameter $\kappa$ differs from another often used convention by a factor of 2,
but has the advantage of keeping intact the projectors $(1\pm \gamma_\mu)/2$.}.  
We use the convention $U_{-j}(\vec{x},x_4) = U_j(\vec{x} -
\hat{j},x_4)^\dagger$, where $\hat{j}$ denotes the unit vector in $j$-direction.
The term $D^{(3)}$ is obtained by shifting the range of the auxiliary sums
running over $n_4$ in the last two terms of (\ref{D11}) by $\beta/2$. 

The terms $D^{(2)}$ and $D^{(4)}$ live on only a single time slice, with 
$D^{(2)}$ given by
\begin{eqnarray}
\hspace*{-5mm} && 
D^{(2)}(x,y) \, = \, \delta_{\vec{x},\vec{y}} \, \delta_{x_4,0} \, 
\delta_{y_4,0}  
\label{D22} \\
&& \hspace{8mm}
- \, \kappa \sum_{j = \pm 1}^{\pm 3} 
\frac{ 1 \mp \gamma_{|j|}}{2} \, U_j(\vec{x},x_4) \, \delta_{\vec{x} + \hat{j},\vec{y}} \, \delta_{x_4,0} \, 
\delta_{y_4,0} \, .  
\nonumber 
\end{eqnarray}
$D^{(4)}$ is obtained from $D^{(2)}$, by setting the time arguments in the
temporal Kronecker deltas to $\beta/2$.

The terms that connect the domains $\Lambda^{(i)}$ have only hops in time
direction, with
\begin{eqnarray}
\hspace*{-5mm} && 
D^{(1,2)}(x,y) \, = \, 
- \, \kappa  
\frac{ 1 - \gamma_4}{2} \, U_4(\vec{x},-1) \, 
\delta_{\vec{x},\vec{y}} \, \delta_{x_4,-1} \, 
\delta_{y_4,0} \, ,  
\nonumber \\
\hspace*{-5mm} && 
D^{(2,1)}(x,y) \, = \, 
- \, \kappa  
\frac{ 1 + \gamma_4}{2} \, U_4(\vec{x},-1)^\dagger \, 
\delta_{\vec{x},\vec{y}} \, \delta_{x_4,0} \, 
\delta_{y_4,-1} \, ,
\nonumber \\
\hspace*{-5mm} &&  \label{D12}
\end{eqnarray}
and the terms $D^{(2,3)}, D^{(3,2)}, D^{(3,4)}$ and $D^{(4,3)}$ are obtained
from $D^{(1,2)}$ and $D^{(2,1)}$ by setting the time arguments according to  the
values listed in (\ref{subdiracs}). 

The terms $D^{(1,4)}$ and $D^{(4,1)}$, which periodically close the lattice in
time direction, have a slightly different form, since here the chemical
potential $\mu$ is coupled and the anti-periodic boundary conditions in time are
implemented. We write the two terms as
\begin{equation}
D^{(4,1)} \, = \, e^{\mu \beta} \, D^{(4,1)}_0 \; ,\; 
D^{(1,4)} \, = \, e^{- \mu \beta} \, D^{(1,4)}_0 \; , \label{mucoupling}
\end{equation}
with the $\mu$-independent pieces 
\begin{eqnarray}
\hspace*{-5mm} && 
D^{(4,1)}_0(x,y) =  
\kappa  
\frac{ 1 \!-\! \gamma_4}{2} U_4(\vec{x},\beta/2) \,  
\delta_{\vec{x},\vec{y}} \, \delta_{x_4,\frac{\beta}{2}} \, 
\delta_{y_4,-\frac{\beta}{2}\!+\!1} ,  
\nonumber \\
\hspace*{-5mm} && 
D^{(1,4)}_0(x,y) = 
\kappa  
\frac{ 1 \!+\! \gamma_4}{2} U_4(\vec{x},\beta/2)^\dagger 
\delta_{\vec{x},\vec{y}} \, \delta_{x_4,-\frac{\beta}{2}\!+\!1} 
\delta_{y_4,\frac{\beta}{2}}.
\nonumber \\
\hspace*{-5mm} &&  \label{D14}
\end{eqnarray} 
Combining all terms $D^{(i)}$ on the individual domains and the connecting
pieces $D^{(i,j)}$, one obtains the standard Wilson Dirac operator
with the chemical potential introduced as a boundary condition. 

For later use we remark that the individual terms obey the following 
$\gamma_5$-hermiticity properties:
\begin{eqnarray}
\hspace*{-5mm} && D^{(i)\;\dagger} \, = \; \gamma_5 \, D^{(i)} \, 
\gamma_5 \; \; , \; \; i = 1,2,3,4, 
\nonumber \\ 
\hspace*{-5mm} && D^{(i,j) \; \dagger} \, = \; \gamma_5 \, D^{(j,i)} \, 
\gamma_5 \; \; , \; \; (i,j) = (1,2),(2,3),(3,4),
\nonumber \\ 
\hspace*{-5mm} && D_0^{(4,1) \; \dagger} \, = \; \gamma_5 \, D^{(1,4)}_0 \, 
\gamma_5 \; .
\label{g5hermiticity}
\end{eqnarray}

We remark at this point, that our approach may be implemented in the same way 
for clover improved Wilson fermions and also staggered fermions. For the latter case a dimensional reduction formula obtained in a different way can be found in \cite{staggeredreduction,hasenfratz}.

\section{Factorization of the determinant}

We now use the partition of the lattice into domains to factorize the fermion
determinant. The basic step is to write the Grassmann integral for the
determinant in terms of the Grassmann fields organized according to the domains
$\Lambda^{(i)}$:
\begin{eqnarray}
\det [D]  & = & \int \prod_{i=1}^4 d \overline{\psi}^{(i)} d \psi^{(i)}
\exp\Big( - \sum_{j=1}^4 \overline{\psi}^{(j)} D^{(j)} \psi^{(j)}
\nonumber \\
&& -  \overline{\psi}^{(1)} D^{(1,2)} \psi^{(2)} 
- \overline{\psi}^{(2)} D^{(2,1)} \psi^{(1)} 
\nonumber \\
&& -  \overline{\psi}^{(2)} D^{(2,3)} \psi^{(3)} 
- \overline{\psi}^{(3)} D^{(3,2)} \psi^{(2)} 
\nonumber \\
&& -  \overline{\psi}^{(3)} D^{(3,4)} \psi^{(4)} 
- \overline{\psi}^{(4)} D^{(4,3)} \psi^{(3)} 
\nonumber \\
&& -  \overline{\psi}^{(4)} D^{(4,1)} \psi^{(1)} 
- \overline{\psi}^{(1)} D^{(1,4)} \psi^{(4)} \Big).
\end{eqnarray}
Here we use vector/matrix notation for all indices, color, Dirac and space-time.
The superscripts attached to the Grassmann variables $\overline{\psi}^{(i)},
\psi^{(i)}$ denote which domains of the lattice these spinors live on. The sum
of all terms in the exponential function reproduces Wilson's fermion action with
chemical potential. 

We now first integrate out the Grassmann variables in the domains 
$\Lambda^{(1)}$ and $\Lambda^{(3)}$, then in domain $\Lambda^{(2)}$ and finally
in $\Lambda^{(4)}$. Solving this chain of Gaussian integrals, one may write the
determinant as
\begin{equation}
\det[D] \; = \; A_0 \, W \; ,
\label{factorform}
\end{equation}
with
\begin{eqnarray} 
A_0 & = & \det\big[ D^{(1)}\big] \, \det\big[ D^{(3)}\big] \, 
\det \big[ \widetilde{D}^{(2)}\big] \, \det\big[ \widetilde{D}^{(4)}\big] ,
\nonumber \\
W
& = &
\det\Big[1 - \widetilde{S}^{(4)} 
\big[e^{\mu \beta} \widetilde{D}^{(4,2)}_1 + \widetilde{D}^{(4,2)}_3\big]  	
\nonumber \\
&&  
\hspace{8mm}\times \, \widetilde{S}^{(2)} 
\big[e^{-\mu \beta} 
\widetilde{D}^{(2,4)}_1 + \widetilde{D}^{(2,4)}_3\big] \Big] .
\label{factordefs}			
\end{eqnarray}
We have introduced the following abbreviations:
\begin{eqnarray}
\hspace*{-3mm} && 
S^{(1)} \, = \, \big(D^{(1)}\big)^{-1} \; , \; \;
S^{(3)} \, = \, \big(D^{(3)}\big)^{-1} \; ,
\nonumber \\
\hspace*{-3mm} &&
\widetilde{D}^{(2)} \, = \; 
D^{(2)} - D^{(2,1)} S^{(1)} D^{(1,2)} - D^{(2,3)} S^{(3)} D^{(3,2)} ,
\nonumber \\
\hspace*{-3mm} && 
\widetilde{D}^{(4)} \, = \; 
D^{(4)} - D^{(4,1)}_0 S^{(1)} D^{(1,4)}_0 - D^{(4,3)} S^{(3)} D^{(3,4)} ,
\nonumber \\
\hspace*{-3mm} &&
\widetilde{S}^{(2)} \, = \, \big(\widetilde{D}^{(2)}\big)^{-1} \; , \; \;
\widetilde{S}^{(4)} \, = \, \big(\widetilde{D}^{(4)}\big)^{-1} \; ,
\nonumber \\
\hspace*{-3mm} &&
\widetilde{D}^{(4,2)}_1 \; = \; D^{(4,1)}_0 S^{(1)} D^{(1,2)} \; ,
\nonumber \\
\hspace*{-3mm} &&
\widetilde{D}^{(4,2)}_3 \; = \; D^{(4,3)} S^{(3)} D^{(3,2)} \; ,
\nonumber \\
\hspace*{-3mm} &&
\widetilde{D}^{(2,4)}_1 \; = \; D^{(2,1)} S^{(1)} D^{(1,4)}_0 \; ,
\nonumber \\
\hspace*{-3mm} &&
\widetilde{D}^{(2,4)}_3 \; = \; D^{(2,3)} S^{(3)} D^{(3,4)} \; .
\label{tildedefs}
\end{eqnarray}
The individual terms have an interesting hierarchical structure: $D^{(1)}$  and
$D^{(3)}$ simply are the Dirac operators on the lattice domains 
$\Lambda^{(1)}$ and $\Lambda^{(3)}$. The corresponding propagators are denoted
by $S^{(1)}$ and $S^{(3)}$.

To leading order the operator $\widetilde{D}^{(2)}$ (and similarly
$\widetilde{D}^{(4)}$) is also the Dirac operator on the corresponding domain
$\Lambda^{(2)}$ (and $\Lambda^{(4)}$ respectively).  However, also two
additional correction terms appear: In the first term the temporal links in
$D^{(2,1)}$ allow a quark to hop into the neighboring domain $\Lambda^{(1)}$.
There it may propagate with the propagator $S^{(1)}$, and in the end is again
transported back into $\Lambda^{(2)}$ by the link terms in $D^{(1,2)}$. The
second correction term is built in the same way, now allowing the quark to
propagate in $\Lambda^{(3)}$. The term $\widetilde{D}^{(4)}$ has exactly the
same structure: It adds to the Dirac operator in $\Lambda^{(4)}$ additional
terms that allow a quark to visit the neighboring domains $\Lambda^{(1)}$ and
$\Lambda^{(3)}$.

A useful way of thinking about the individual contributions is in terms of
paths: The propagators $S^{(1)}$ and $S^{(3)}$ may be viewed as collections of
paths that connect any point in $\Lambda^{(1)}$ ($\Lambda^{(3)}$) with any other
point. Along these paths ordered products of the gauge links $U_\mu(x)$ appear,
such that the individual contributions are gauge covariant. The correction terms
in $\widetilde{D}^{(2)}$ and $\widetilde{D}^{(4)}$ thus are paths in the
neighboring domains $\Lambda^{(1)}$ and $\Lambda^{(3)}$ which are connected  to
the time slices $\Lambda^{(2)}$ and $\Lambda^{(4)}$ through temporal hops. 

It is important to note, that the four determinants in the factor $A_0$
(\ref{factordefs}) do not depend on the chemical potential $\mu$. The only term
where the chemical potential could appear in $A_0$ is the first correction term in
$\widetilde{D}^{(4)}$. However, since there both $D^{(4,1)}$ and $D^{(1,4)}$
appear, the factors $\exp(\pm \mu \beta)$ cancel, and only the $\mu$-independent
terms $D^{(4,1)}_0$ and $D^{(1,4)}_0$ remain. As a consequence one expects that
the four determinants are real. For $\det[ D^{(1)}]$ and $\det[D^{(3)}]$ this
follows from the $\gamma_5$-hermiticity of $D^{(1)}$ and $D^{(3)}$ (see
Eq.~(\ref{g5hermiticity})). Also the corrected Dirac operators
$\widetilde{D}^{(2)}$ and $\widetilde{D}^{(4)}$ are $\gamma_5$-hermitian, as can
be established with the help of the relations in (\ref{g5hermiticity}). The
corresponding determinants thus also are real.   

We conclude, that the complex phase must be contained in the factor $W$ defined
in (\ref{factordefs}),  where we explicitly see terms that do depend on the
chemical potential $\mu$.  In the factor $W$  new types of matrices appear, such
as the matrix $\widetilde{D}^{(4,2)}_1$: This matrix connects the domains
$\Lambda^{(4)}$ and $\Lambda^{(2)}$ by first hopping from $\Lambda^{(4)}$ into
$\Lambda^{(1)}$ with the temporal hops in $D^{(4,1)}_0$, subsequently
transporting a quark with $S^{(1)}$, and finally connecting to $\Lambda^{(2)}$
with the hops in $D^{(1,2)}$. Thus $\widetilde{D}^{(4,2)}_1$ 
may be considered as
a {\it large forward hopping term} from $\Lambda^{(4)}$ into $\Lambda^{(2)}$
going through $\Lambda^{(1)}$. In the same way  $\widetilde{D}^{(4,2)}_3$ can be
viewed as a {\it large backward hopping term} from $\Lambda^{(4)}$ into
$\Lambda^{(2)}$ going through $\Lambda^{(3)}$. The other two new terms
$\widetilde{D}^{(2,4)}_3$ and $\widetilde{D}^{(2,4)}_1$ have a similar
interpretation as large forward and backward hopping terms.

We stress at this point, that the paths which the large hops consist of have a
length of $\beta/2$ or longer, i.e., they consist of at least $\beta/2$
individual steps. Let us discuss this property for the example of $D^{(4,2)}_1$:
According to (\ref{tildedefs}) this term is given by the the product 
$D^{(4,1)}_0 S^{(1)} D^{(1,2)}$. The first and the last factor each contribute
one temporal step. The propagator $S^{(1)}$ is a collection of arbitrary paths,
but here we only need to consider the paths that extend from the timeslice with
$x_4 = -\beta/2+1$ to the timeslice at $x_4 = - 1$, since otherwise the terms
$D^{(4,1)}_0$ and $D^{(1,2)}$ could not attach to $S^{(1)}$ (compare
Fig.~\ref{latticeplot}). The paths that connect the two boundaries of
$\Lambda^{(1)}$ have at least a length of $\beta/2-2$ steps, which establishes,
that $D^{(4,2)}_1$ indeed consists of paths which have at least the length
$\beta/2$, and the same holds for $D^{(4,2)}_3$, $D^{(2,4)}_1$ and
$D^{(2,4)}_3$. 

Finally we remark that also the large hopping terms are related to each other
through $\gamma_5$-hermiticity:
\begin{eqnarray}
D^{(2,4) \, \dagger}_1 & = & \gamma_5 \, D^{(4,2)}_1 \, \gamma_5 \; ,
\nonumber \\
D^{(2,4) \, \dagger}_3 & = & \gamma_5 \, D^{(4,2)}_3 \, \gamma_5 \; .
\label{largehopg5} 
\end{eqnarray}

Effectively the factor $W$ has the structure  of a determinant with only two
timeslices given by the domains $\Lambda^{(2)}$ and $\Lambda^{(4)}$. With the
large hops  $\widetilde{D}^{(2,4)}_1$, $\widetilde{D}^{(2,4)}_3$,
$\widetilde{D}^{(4,2)}_1$ and $\widetilde{D}^{(4,2)}_3$ we can hop between these
timeslices. The terms $S^{(4)}$ and $S^{(2)}$ that appear in $W$ between the
large hops allow for a local propagation within $\Lambda^{(2)}$ and
$\Lambda^{(4)}$.

\section{Decomposition into winding sectors}

The factor $W$ of (\ref{factordefs}) may be expanded with the trace-log formula
for determinants. For that purpose we introduce the abbreviations
\begin{eqnarray}
H_0 \;\;& = & \widetilde{S}^{(4)} D^{(4,2)}_1 \widetilde{S}^{(2)} D^{(2,4)}_1 + 
\widetilde{S}^{(4)} D^{(4,2)}_3 \widetilde{S}^{(2)} D^{(2,4)}_3 ,
\nonumber \\
H_{+1} & = & \widetilde{S}^{(4)} D^{(4,2)}_1 \widetilde{S}^{(2)} D^{(2,4)}_3 \, ,
\nonumber \\
H_{-1} & = & \widetilde{S}^{(4)} D^{(4,2)}_3 \widetilde{S}^{(2)} D^{(2,4)}_1 \, .
\label{hdefs}
\end{eqnarray}
These terms may all be viewed as paths that connect two points in
$\Lambda^{(4)}$. In $H_0$ these paths visit $\Lambda^{(2)}$ by either going
through $\Lambda^{(1)}$ or $\Lambda^{(3)}$ and then coming back through the same
domain. The paths in $H_{\pm1}$ visit all four domains, i.e, they wind once
around the lattice. In $H_{+1}$ they wind in forward direction, while in
$H_{-1}$ they run backwards. According to the above discussion, in all three
terms $H_0$,$H_{\pm1}$ the paths have a minimum length of $\beta$ steps.   

With the help of the definitions in (\ref{hdefs}) and the trace-log formula for
determinants we can now write $W$ as
\begin{eqnarray}
\hspace*{-5mm}
&& W \, = \, \det 
\big[ 1 - H_0 - e^{\mu \beta} H_{+1} - e^{-\mu \beta} H_{-1} \big] 
\label{tracelog} \\ 
\hspace*{-5mm}&& 
= \, \exp\!\left(\!- \!\sum_{n=1}^\infty \frac{1}{n} \mbox{Tr} \;
\big[ H_0 + e^{\mu \beta} H_{+1} + e^{-\mu \beta} H_{-1} \big]^n \!\right)
\nonumber \\ 
\hspace*{-5mm}&& 
= \, \exp\!\left(\!-
\!\sum_{q \in \mathds{Z}} e^{\mu \beta q} \sum_{n=1}^{\infty} 
\frac{1}{n} 
\hspace{-5mm} \sum_{\hspace{5mm} k_1+ ...  + k_n = q} \hspace{-7mm} 
\mbox{Tr} \;
\big[ H_{k_1} H_{k_2}  ...   H_{k_n} \big] \!\! \right)\!\!.
\nonumber 
\end{eqnarray}
In the final step we have rewritten the $n$-th powers by introducing  two
auxiliary sums: The first sum runs over an integer variable $q$ which below we
will identify as the winding number of the individual loop contributions.  The
second, restricted sum runs over $n$ variables $k_i \in \{-1,0,+1\}$. 

\begin{table*}[t!]
\begin{center}
\begin{tabular}{c|ccccccc}
$m$ & $A_0$ & 
$|\det[D]|_{\mu=0}$ \;&\; $|W|_{\mu=0}$ \;&\; 
$|\det[D]|_{\mu = 0.1}$ \;&\; $|W|_{\mu = 0.1}$ \;&\; 
$|\det[D]|_{\mu = 0.2}$ \;&\; $|W|_{\mu = 0.2}$ \\
\hline \\
0.05 \;&\; $0.724\!\times\!10^{47}$ \;&\; 
$0.232\!\times\!10^{48}$ \;&\; 3.202 \;&\; 
$0.223\!\times\!10^{48}$ \;&\; 3.085 \;&\;
$0.199\!\times\!10^{48}$ \;&\; 2.743 \\

0.10 \;&\; $0.179\!\times\!10^{45}$ \;&\; 
$0.509\!\times\!10^{45}$ \;&\; 2.838 \;&\; 
$0.492\!\times\!10^{45}$ \;&\; 2.740 \;&\;
$0.440\!\times\!10^{45}$ \;&\; 2.453 \\

0.20 \;&\; $0.377\!\times\!10^{40}$ \;&\; 
$0.875\!\times\!10^{40}$ \;&\; 2.319 \;&\;
$0.848\!\times\!10^{40}$ \;&\; 2.249 \;&\;
$0.769\!\times\!10^{40}$ \;&\; 2.039 \\
\end{tabular}
\end{center}
\caption{Values of $A_0$, $|\det[D]|$ and $|W|$ for three different quark masses, 
$m = 0.05,0.1,0.2$, and three different values of the chemical potential, 
$\mu = 0, 0.1, 0.2$ for a single configuration.}
\label{factortable}
\end{table*}

Due to the trace the individual terms $\mbox{Tr} \, [ H_{k_1} H_{k_2}  ...  
H_{k_n}]$ correspond to closed loops built from the paths in the $H_{k_i}$.
Since $H_{+1}$ and $H_{-1}$ contribute a path that winds once in forward
(backward) direction and $H_0$ does not wind at all, the sum $q = k_1 + k_2 + \,
... \, k_n$ is the winding number of the loop. It is obvious from the last
expression in (\ref{tracelog}) that the chemical potential couples to the
winding number as expected. 

From (\ref{tracelog}) follows the factorization of $W$:
\begin{equation}
W \, = \, \prod_{q \in \mathds{Z}} W^{(q)} \; ,
\label{Wfactor}
\end{equation}
where the factors for the individual winding sectors are given by 
\begin{equation}
W^{(q)}  = \; 
\exp\!\left(\!- e^{\mu q \beta} \sum_{n=1}^{\infty} 
\frac{1}{n} 
\hspace{-6mm} \sum_{\hspace{5mm} k_1+ ...  + k_n = q} \hspace{-8mm} 
\mbox{Tr} \;
\big[ H_{k_1} H_{k_2}  ...   H_{k_n} \big] \!\! \right)\!\!.
\label{wcomponents}
\end{equation}

We remark that there is a relation between the sectors of opposite winding
number $q$,
\begin{equation}
W^{(-q)} (\mu ) \; = \; W^{(q)} (- \mu)^* \; ,
\label{qrelation}
\end{equation}
where the asterisk denotes complex conjugation. This follows again from the
$\gamma_5$-hermiticity relations (\ref{g5hermiticity}) which can be used to
establish
\begin{equation}
\mbox{Tr} \;
\big[ H_{k_1} H_{k_2}  ...   H_{k_n} \big] \; = \; 
\mbox{Tr} \;
\big[ H_{k_1^\prime} H_{k_2^\prime}  ...   H_{k_n^\prime} \big]^* \, ,
\label{traceconjugate}
\end{equation}
where  $k_1^\prime = - k_n$, $k_2^\prime = - k_{n-1}$, ... $k_n^\prime = -
k_{1}$. This result immediately leads to (\ref{qrelation}).

For zero chemical potential the relation (\ref{qrelation}) reduces to 
\begin{equation} 
W^{(-q)} (\mu = 0) \; = \; W^{(q)} (\mu = 0)^* \; ,
\label{qreality} 
\end{equation} 
which ensures the well known fact that the determinant is real for vanishing
chemical potential ($W^{(0)}$  is real anyway).  It is quite interesting to
note, that due to the relation (\ref{qreality}) at $\mu =0$ the contributions
from the non-trivial sectors with $q \neq 0$ are positive, since $W^{(-q)} (0)
\, W^{(q)} (0) =   |W^{(q)} (0)|^2$. Thus a possible negative sign of the $\mu =
0$ determinant must come from the non-winding contributions $A_0 W^{(0)}$.  For
$\mu \neq 0$ the relation (\ref{qreality}) is violated due to the different
weight factors $\exp(\pm \mu q \beta)$, and the determinant becomes complex.

Let us finally discuss an important aspect of our result: We have noted above
that the paths in the terms $H_{k_i}$ consist of at least $\beta$ steps. Thus a
contribution  $\mbox{Tr} \, [ H_{k_1} H_{k_2}  ...   H_{k_n}]$ consists of
closed loops which have at least a length of $n \, \beta$. Long loops are
suppressed exponentially in $n$, due to the fluctuations of the gauge links and
the factors of $\kappa$. Obviously the sum over $n$ corresponds to a generalized
hopping expansion.

\section{Numerical analysis of the factorization formula}  

Various aspects of our determinant factorization can be analyzed
numerically. In particular we study the behavior of the factors $A_0$
and $W$ as a function of the mass parameter $m$ and the chemical potential
$\mu$. For our exploratory analysis we use quenched \footnote{It has to be
remarked, that the quenched ensemble used here only serves to test properties of
the analytical results obtained in this article. Configurations from a fully dynamical
ensemble would certainly give quantitatively different results.} gauge
configurations on lattices of size $8^3 \times 4$. The configurations were
generated with the L\"uscher-Weisz gauge action \cite{LuWeact} and at the value
of $\beta = 7.60$ which we use here, the lattice spacing was determined
\cite{scale} to be $a= 0.193$ fm. This gives rise to a cutoff of roughly 1 GeV
and corresponds to a temperature of 254 MeV. 

As a first test we programmed both sides of the 
exact relation (\ref{factorform}) in order to test 
the correct implementation of the terms in (\ref{factordefs}). 
We found agreement to machine precision.    

In the actual analysis of the properties of the factorization formula we begin
with evaluating the factors $A_0$ and $W$  (see Eq.\ (\ref{factordefs})) as a
function of mass $m$ and chemical potential $\mu$. We refer to $A_0$ as the {\sl
bulk factor}, while $W$ is the {\sl winding term} of the determinant. For the
numerical study we consider the mass values $m = 0.05, 0.1, 0.2$, and scan the
chemical potential  in the interval $\mu \in  [0.0, 1.0]$ in steps of 0.01.
Concerning the chemical potential we remark, that once the matrices $A_0, H_0,
H_{\pm1}$ are evaluated, it is very cheap to compute the  $\mu$-dependent
winding  term $W$ for many values of $\mu$. The reason is obvious from the
definition of $W$ in the first line of (\ref{tracelog}), which shows that the
determinant one has to evaluate is only over the indices of a single time slice.
This dimensional reduction speeds up the numerical evaluation of the determinant
considerably. 

To give an idea of the relative size of the different quantities, we show in
Table 1 the values of bulk factor $A_0$ together with the absolute values
$|\det[D]|$ and $|W|$ for three different masses $m = 0.05,0.1$ and 0.2, for a
single gauge configuration. Note that $A_0$ is independent of $\mu$, while
$|\det[D]|$ and $|W|$ do depend on $\mu$, and we thus give the latter two factors for three
different values of the chemical potential, $\mu = 0.0,0.1,0.2$.  

What is immediately obvious from the table, is the rather different scale of the
two factors $A_0$ and $W$, which according to (\ref{factorform}) constitute the
determinant, $\det[D] = A_0 W$. While the bulk factor $A_0$ is large of order
${\cal O}(10^{40})$ -- ${\cal O}(10^{47})$, depending on the mass parameter $m$, the winding term $W$ is of order 1 (at
least for the values of $\mu$ in the table). Furthermore $A_0$ strongly
depends on the mass, while $W$ shows only a weak dependence. We thus conclude
that the real valued bulk factor $A_0$ contributes the mass-dependent scale of
the determinant, while $W$ brings in the $\mu$-dependence and the complex
phase. Thus, besides separating the $\mu$-dependence, our decomposition 
provides a natural factorization into the relevant scales for the bulk behavior 
and the response to the chemical potential.

\begin{figure}[t]
\begin{center}
\includegraphics[width=7.38cm,clip]{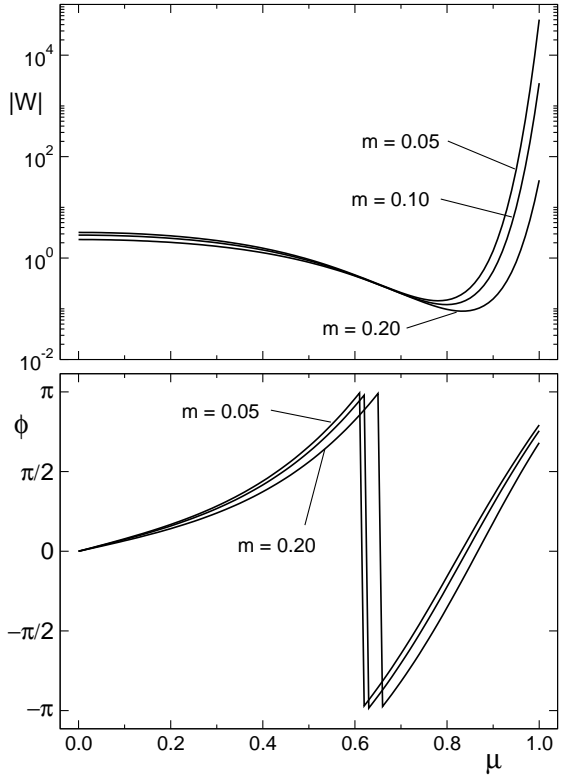}
\end{center} 
\caption{Absolute value (top plot) and phase (bottom) of 
the winding term $W$ as a function of $\mu$ for three different 
masses on a single configuration.
\label{W_vs_mu}}
\end{figure}

In Fig.~\ref{W_vs_mu} we now analyze how the winding term $W$ changes with the
chemical potential $\mu$. In the top plot we show the absolute value $|W|$ as a
function of $\mu$, while in the bottom plot the phase $\phi$ is displayed. Note
that since $A_0$ is real, the phase $\phi$ of $W$ is also the phase of $\det[D]$
(up to a possible minus sign that could appear for exceptional configurations). 
The data in Fig.~\ref{W_vs_mu} are for a single configuration which we analyzed
for three different values of the quark mass, $m = 0.05,0.1,0.2$.

The plots show a non-trivial dependence of $W$ on $\mu$. Starting out with a weak $\mu$-dependence for small $\mu$, $W$ changes considerably as $\mu$ gets close to the cutoff at $\mu = 1$. With increasing $\mu$
the phase $\phi$ runs twice through all of the
interval $[-\pi,\pi]$ (note that when exceeding $\pi$ we project the phase back
into $[-\pi,\pi]$). What is surprising is that the mass dependence for both,
$|W|$ and the phase $\phi$ is rather small. This supports the above sketched
picture of $A_0$ carrying most of the overall scale and mass dependence.

We remark that we repeated the analysis of Fig.~\ref{W_vs_mu} for several
configurations and found that the scale aspects of the behavior we describe in
the last paragraph are universal: $A_0$ carries most of the overall scale of the
determinant and also of the mass dependence. The $\mu$-dependence in $W$ shows a
much smaller variation. Other aspects are specific for individual
configurations, in particular how quickly and with which orientation the phase
$\phi$ is run through as a function of $\mu$. Also the details of the behavior
of  $|W|$ as a function of $\mu$ may vary for different configurations. 

\begin{figure}[t]
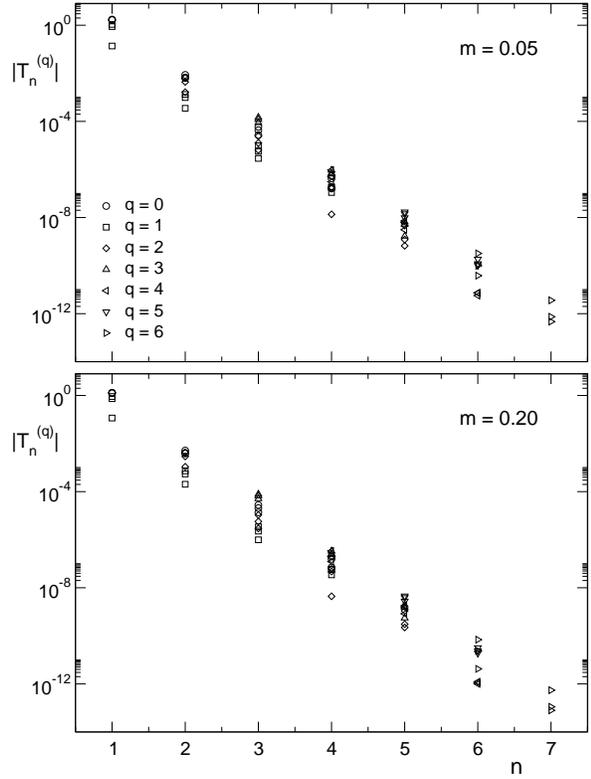

\begin{center}
\includegraphics[width=7.7cm,clip]{Tqn_8x4_m0.05.eps} 
\vskip0.8mm
\includegraphics[width=7.7cm,clip]{Tqn_8x4_m0.2.eps} 
\end{center}
\caption{The absolute values $|T_n^{(q)}|$ for various values of $q$ as a
function of $n$. We use two different masses, $m = 0.05$ in the plot at the top,
and $m = 0.20$ at the bottom. We superimpose the data from three different gauge
configurations.
\label{Tq_plot}}
\end{figure}

Let us now analyze the buildup of the contributions in the individual winding
sectors. For that purpose we write the factor $W^{(q)}$ for winding number $q$
as 
\begin{equation}
W^{(q)} \; = \; \exp  \left( - e^{\mu q \beta} \, T^{(q)} \right) \; ,
\end{equation}
where we define
\begin{eqnarray}
T^{(q)} & = & \sum_{n=1}^\infty T^{(q)}_n \; ,
\label{tqdefs} \\
T^{(q)}_n & = & \frac{1}{n} \sum_{k_1+ ...  + k_n = q}    \mbox{Tr} \, [ H_{k_1} H_{k_2}  ...   H_{k_n}] \; .
\nonumber
\end{eqnarray}
It is straightforward to evaluate the coefficients $T^{(q)}_n$ and for
illustration purposes we list a few terms:
\begin{eqnarray}
\hspace*{-5mm} && 
T_1^{(0)} = \mbox{Tr} \, H_0 , 
\\
\hspace*{-5mm} &&
T_2^{(0)} = \mbox{Tr} \, H_{+1} H_{-1}  + \frac{1}{2} \mbox{Tr} \, H_0^2 ,	
\nonumber \\
\hspace*{-5mm} &&
T_3^{(0)} = \mbox{Tr} \, H_{+1} H_0 H_{-1}  + \mbox{Tr} \, H_{+1} H_{-1} H_0  + 
\frac{1}{3} \mbox{Tr} \, H_0^3 .
\nonumber
\end{eqnarray} 	
\begin{eqnarray}
\hspace*{-26mm} && 
T_1^{(1)} = \mbox{Tr} \, H_{+1} \; \; , \; \; 
T_2^{(1)} = \mbox{Tr} \, H_{+1} H_{0} ,
\\
\hspace*{-26mm} &&
T_3^{(1)} = \mbox{Tr} \, H_{+1}^2 H_{-1}  + \mbox{Tr} \, H_{+1} H_0^2 .
\nonumber
\end{eqnarray} 					
\begin{equation} 
\hspace*{-1mm}
T_1^{(2)} = 0 \; , \;  
T_2^{(2)} = \frac{1}{2} \mbox{Tr} \, H_{+1}^2 \; , \; 
T_3^{(2)} = \mbox{Tr} \, H_{+1}^2 H_{0} . \hspace{5mm}
\end{equation} 					
\begin{equation} 
\hspace*{-19mm}
T_1^{(3)} = 0 \; , \; 
T_2^{(3)} = 0 \; , \; 
T_3^{(3)} = \frac{1}{3} \mbox{Tr} \, H_{+1}^3 .  
\end{equation}
From $q = k_1 + k_2 + ... \, k_n$ follows $T^{(q)}_n = 0$ for $q > n$. 
Note that due to (\ref{traceconjugate}) the coefficients obey
\begin{equation}
T^{(0)}_n \, \in \, \mathds{R} \; \; , \; \; 
T^{(-q)}_n \, = \, T^{(q) \; *}_n \; .
\label{qflipproperties}
\end{equation}
We stress again that the minimal length of the loops in a coefficient 
$T_n^{(q)}$ is $n \beta$, such that we expect the $T_n^{(q)}$ to decrease
exponentially with $n$.

The evaluation of the $T^{(q)}_n$ may numerically be implemented efficiently by
pre-calculating often used combinations such a $H_{\pm 1}^2$, $H_{\pm1} H_0$
etc. Note that all these terms only live on the single time slice
$\Lambda^{(4)}$. Thus the storage requirements are modest and the matrix
multiplications and traces needed for the $T^{(q)}_n$ run only over the indices
in a single time slice. 

In Fig.~\ref{Tq_plot} we show the absolute values $|T^{(q)}_n|$ for various $q =
0,1 ... 7$ as functions of $n$. We compare two values of the mass parameter,
using $m = 0.05$ in the top plot and $m = 0.20$ at the bottom. To illustrate the
fluctuation of the absolute values we superimpose the data points from three
different gauge configurations.    The plots nicely display the expected
exponential decay with increasing $n$,  with the data points for different $q$
and different configurations all being close to a straight line.  It is
interesting to note, that although between the top and the bottom plots the mass
was increased by a factor of 4, the slope of this line changes only very little.
This is not a surprise, since the mass $m$ enters through powers  of $\kappa =
1/(4+m)$ which for the range of masses considered in the plots is only a slowly
varying function.  Also for the phase of the $T^{(q)}_n$ we find only a very
small mass dependence. The weak mass dependence of the $T^{(q)}_n$  then in turn
implies the weak mass dependence of $W$ which we have observed in 
Fig.~\ref{W_vs_mu} and Table I.

\section{Applications of the factorization formula}

The winding decomposition formula can be applied to the determination of the
canonical partition functions for a fixed quark or baryon number.  The grand
canonical partition function $Z_{GC}(\mu)$ at chemical potential $\mu$ is
obtained by integrating the determinant over the gauge fields,
\begin{equation}
Z_{GC}(\mu) \; = \; \int {\cal D}[U] \, e^{-S_G[U]} \, \det[D(\mu)] \; ,
\label{grandcanonical}
\end{equation}
where $S_G[U]$ is the gauge action, and for simplicity we discuss the case of
only one quark flavor. 

The grand canonical partition function may be written as a sum over canonical
partition functions $Z^{(Q)}$ with a fixed quark number $Q$, 
\begin{equation}
Z_{GC}(\mu)  \; = \; \sum_{Q} \, e^{\mu Q \beta} Z^{(Q)}_C \; =
\; \sum_{B} \, e^{3 \mu B \beta} \, Z^{(3B)}_C \; ,
\end{equation} where in the second step we have made explicit the fact that due
to the center symmetry only the canonical partition functions $Z^{(3B)}_C$,
where the quark number is a multiple of 3, are non-vanishing. $B$ is the baryon
number. We remark, that for expectation values also quark sectors where $Q$ is
not a multiple of 3 contribute, and furthermore stress that the center symmetry may be broken at the phase transition.

Our winding decomposition formula may now be used to determine the canonical
partition functions in various effective ways.

\subsection{Projected determinants from the Fourier transformation method}

The canonical partition functions may, e.g., be determined from a Fourier transformation with respect to an imaginary chemical potential $\mu = i\varphi/\beta$:
\begin{equation}
Z_C^{(Q)} \; = \; \int {\cal D}[U] \, e^{-S_G[U]} \, D^{(Q)} \; ,
\label{canonical}
\end{equation}
where we define the {\sl projected determinants} $D^{(Q)}$,
\begin{equation}
D^{(Q)} \; = \; \int_{-\pi}^\pi \frac{d \varphi}{2\pi} \, 
e^{-i Q \varphi} \, \det[D(\mu = i\varphi/\beta)] \; .
\label{Dk_def}
\end{equation}
Although the determination of the $D^{(Q)}$ through Fourier transformation
assumes the elegant form of Eq.~(\ref{Dk_def}), it is expensive in an actual
implementation. The $\varphi$ integral has to be done numerically and experience
shows that a large number of $\varphi$-values is needed for an accurate
approximation of the $\varphi$-integral. Thus the determinant $\det[D(\mu =
i\varphi/\beta)]$ needs to be evaluated many times for different $\varphi$.

With the factorization formula this procedure may be accelerated considerably. 
Using (\ref{factorform}), (\ref{tracelog}) we find
\begin{eqnarray}
D^{(Q)} & = & \frac{A_0}{2\pi} \! \!\int_{-\pi}^\pi \!\!\!\! d \varphi \, 
e^{-i Q \varphi} 
\label{fourierfactor}
\\
&& \quad \times \det[1 - H_0 - e^{i \varphi} H_{+1} - e^{-i \varphi} H_{-1}].
\nonumber
\end{eqnarray}
Also here a determinant is integrated over $\varphi$ and thus has to be
evaluated at many values of $\varphi$. However, the determinant in
(\ref{fourierfactor}) is for a considerably smaller matrix, since  $H_0, H_{\pm
1}$ live on only a single timeslice. The strategy thus is to pre-calculate $H_0,
H_{\pm 1}$ and $A_0$ and to work with (\ref{fourierfactor}) instead of
(\ref{Dk_def}). The dimension of the problem is reduced by a factor of $\beta$,
the number of timeslices. Since the cost for the exact evaluation of a
determinant is proportional to the third power of the matrix size, one gains a
factor of $\beta^3$, which for typical lattice sizes is an improvement of 
${\cal O}(100)$.  The increased efficiency due to the dimensional reduction allows for a considerably larger
number of values $\varphi$ used in the numerical integration, such that the
accuracy of the results for the projected determinants improves.  

\subsection{Projected determinants from a power series expansion}

In addition to the Fourier transformation technique, the equations derived in
Section III allow also for an efficient determination of the projected
determinants $D^{(Q)}$ via a power series expansion. From 
(\ref{grandcanonical}) -- (\ref{canonical}) follows the fugacity expansion
\begin{equation}
\det[D(\mu)] \; = \; \sum_{Q} \, e^{\mu Q \beta} \, D^{(Q)} \; ,
\end{equation}
which shows that the $D{(Q)}$ may be obtained as the coefficients of a series 
expansion of $\det[D(\mu)]$. 

Using the notation of the last two sections we can write the determinant as
\begin{eqnarray}
&& \det[D(\mu)] \, = \,  A_0 W^{(0)} 
\label{Tqdeterminant} 
\\
&& \hspace{15mm}\times \, 
\prod_{q=1}^\infty  
\exp\left( -e^{\mu q \beta} T^{(q)} - e^{-\mu q \beta} T^{(q) *} \right)  .
\nonumber
\end{eqnarray}
Expanding the exponential functions and ordering the terms with respect to the
powers of $e^{\pm \mu \beta}$ we can read off the projected determinants $D^{(Q)}$
\begin{eqnarray}
D^{(0)} & = & 
A_0 W_0 \!\left( 1 + \sum_{q=1}^\infty |T^{(q)}|^2 + \frac{1}{4} 
\sum_{q=1}^\infty |T^{(q)}|^4 \, ...  \right),
\nonumber \\
D^{(1)} & = & 
A_0 W_0 \! \left( - T^{(1)} + \sum_{q=1}^\infty T^{(q+1)} T^{(q) *}  
\, + \, ...  \right),
\nonumber \\
D^{(2)} & = & 
A_0 W_0 \! \left( - T^{(2)} + \sum_{q=1}^\infty T^{(q+2)} T^{(q) *}  
\, + \, ...  \right),
\nonumber \\
&& \mbox{et cetera} \; .
\label{Dk_expansion}
\end{eqnarray}
Again we can obtain the $D^{(Q)}$ for negative $Q$ via complex conjugation, 
$D^{(-Q)} = D^{(Q) *}$. Note that a term $T^{(q)}$ consists of loops that wind
at least $n = q$ times and thus according to the numerical analysis higher terms
in (\ref{Dk_expansion}) are exponentially suppressed, and the sums may be
truncated after a few leading terms. We remark that the expansion which we have denoted here for a single flavor may easily be generated to an arbitrary number of flavors.

\subsection{A simple truncation scheme}

Having worked out the evaluation of the contributions in the individual winding
sectors we now address the question of a suitable truncation scheme. Obviously
the $T^{(q)}$ are given by infinite sums (see Eq.~(\ref{tqdefs}))  which cannot
be summed in closed form. However, the exponential decrease of the individual
terms $T^{(q)}_n$, which was illustrated in Fig.~\ref{Tq_plot}, suggests 
that a truncation of the sums might lead to a reasonable approximation scheme. We
experimented with truncated contributions $\widehat{T}^{(q)}$ defined as
\begin{eqnarray}
\widehat{T}^{(0)} & = & \sum_{n=1}^{1 + \Delta} T_n^{(0)} \; ,
\label{truncation} \\
\widehat{T}^{(q)} & = & \sum_{n=|q|}^{|q| + \Delta} T_n^{(q)} \; \;\;\;
\mbox{for} \; \; q \neq 0 \; .
\nonumber
\end{eqnarray}
The parameter $\Delta$, which is a positive integer, controls the approximation
and in the limit $\Delta \rightarrow \infty$ the exact result is obtained. Since
the $T^{(q)}_n$ decay exponentially in $n$, it may be expected that already a
small value of $\Delta$ might give a good approximation. In fact 
Fig.~\ref{Tq_plot} shows that for this case subsequent terms are suppressed by a
factor of ${\cal O}(100)$. 

A particularly simple expansion is obtained for $\Delta = 1$. In this case one
finds
\begin{eqnarray}
\hspace*{-7mm} \widehat{T}^{(0)} 
\!&\! = \!&\! \mbox{Tr} \, [H_0] + \frac{1}{2} \, 
\mbox{Tr} \, [(H_0)^2] + \mbox{Tr} \, [H_{+1} H_{-1} ] \; ,
\label{delta1_scheme} \\
\hspace*{-7mm}\widehat{T}^{(q)} \!&\! = \!&\! \frac{1}{q} \, 
\mbox{Tr} \, [(H_{+1})^q] + \mbox{Tr} \, [(H_{+1})^{q} H_{0} ] 
\, , \; (q > 0) .
\nonumber
\end{eqnarray}
Each term (except for $q=0$) consists of only two traces. Furthermore the
arguments of the traces have a particularly simple form which allows for a
efficient recursive implementation of the needed matrix products. 

In order to test the quality of the truncation at $\Delta = 1$, we evaluated the
determinant according to (\ref{Tqdeterminant}), replaced the $T^{(q)}$ by the
$\widehat{T}^{(q)}$ as given in (\ref{delta1_scheme}), and took into account all
terms with $q = 0,1 \, ... \, 6$.  

\begin{figure}[t]
\begin{center}
\includegraphics[width=7.5cm,clip]{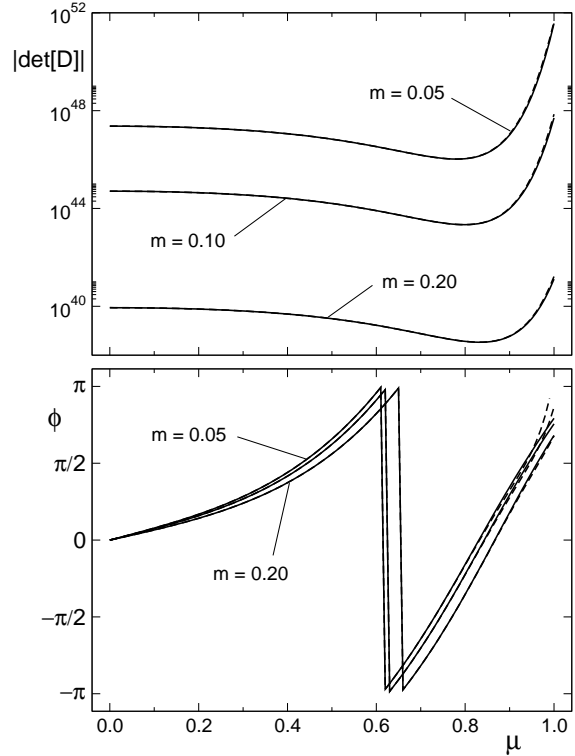}
\end{center} 
\caption{Absolute value (top plot) and phase $\phi$ (bottom) of the determinant
$\det[D]$ as a function of $\mu$. For an individual gauge configuration and
three different masses we compare the exact result (full curves) to the
approximation discussed in Section VI.C.~for $\Delta = 1$ (dashed curves).}
\label{delta1_plot}
\end{figure}

In Fig.~\ref{delta1_plot} we compare the outcome of the truncated determinant
(dashed curves) to the exact result (full curves). The comparison is done on a
single gauge configuration for three different masses. The plot shows that
already the $\Delta = 1$ approximation works surprisingly well. For most of the
$\mu$-range the approximation falls exactly on top of the exact curve and only
when $\mu$ reaches the cutoff at $\mu = 1$, deviations of the approximation 
(dashed curve) become visible. The onset
of these deviations may be pushed to even higher $\mu$ by including larger $q$
in (\ref{Tqdeterminant}), while still sticking to the $\Delta =1$ approximation.
For the quenched example studied here this is not necessary, since the deviation
occurs only close to the cutoff. 

We repeated this analysis for several gauge configurations, and found the same
quality of approximation throughout. Thus we conclude that the truncation
(\ref{truncation}) together with the series expansion of Section VI.B.\ provides
a powerful method for computing the projected determinants $D^{(Q)}$. 

\section{Concluding remarks}

In this paper we have revisited the problem of finding the fermion determinant
in a sector with a fixed quark number, i.e., projecting to a fixed winding
number of the closed loops the determinant consists of. In our approach we use a
partition of the lattice into four domains and the determinant is expressed in
terms of sub-determinants and propagators on the domains. The determinant can be
organized in such a way that the winding numbers of the individual contributions
can be accessed efficiently. 

Our result allows for two different new approaches for computing the projected
determinant at a fixed winding number: The factorization of the determinant
dimensionally reduces the piece which depends on the chemical potential to a
single time slice. Thus the evaluation of the determinant for different values
of the imaginary potential $\varphi$ becomes considerably cheaper, and many more
$\varphi$-points can be used in the numerical Fourier transformation to the
determinant at fixed quark number. A second more direct approach is a power
series expansion of the determinant, where the terms at fixed winding number
appear as expansion coefficients. We propose a suitable truncation scheme for
this expansion which we analyzed numerically. 

The method of a partition into domains may be used also for other applications.
In particular it exists also for propagators and allows for the analysis of the
winding numbers of the paths the propagators 
consist of. This might be useful for the
recently proposed dual chiral condensate \cite{dualcond}, where the Fourier
transform approach is used to define observables that are sensitive to chiral
symmetry breaking as well as confinement. The domain factorization of therm 
propagator might also be combined with a pseudofermion representation of the
fermion determinant and so open another approach to lattice QCD with chemical
potential.  
\\
\\									
{\bf Acknowledgments:} We thank Falk Bruckmann, Philippe de Forcrand, Christian
Lang and Keh-Fei Liu for interesting and helpful discussions. This work is
supported by the Austrian FWF doctoral college DK W1203. C.G.~thanks the Galileo-Galilei
Institute for Theoretical Physics for the hospitality, and the INFN for partial 
support during the completion of this work.

\clearpage
\end{document}